# Improved Particle Confinement with Resonant Magnetic Perturbations in DIII-D Tokamak H-mode Plasmas

N.C. Logan,[1,*] Q. Hu,[2] C. Paz-Soldan,[3,†] R. Nazikian,[2,‡] T. Rhodes,[4] T. Wilks,[5]
S. Munaretto,[3,§] A. Bortolon,[2] F. Laggner,[2] F. Scotti,[1] R. Hong,[4] and H. Wang[3]

[1] *Lawrence Livermore National Laboratory, Livermore, California 94550, USA*
[2] *Princeton Plasma Physics Laboratory, Princeton, New Jersey 08540, USA*
[3] *General Atomics, San Diego, California 92186, USA*
[4] *University of California, Los Angeles, Los Angeles, California 90095, USA*
[5] *Massachusetts Institute of Technology, Cambridge, Massachusetts 02139, USA*

Experiments on the DIII-D tokamak have identified a novel regime in which applied resonant magnetic perturbations (RMPs) increase the particle confinement and overall performance. This work details a robust range of counter-current rotation over which RMPs cause this density pump-in effect for high confinement (H-mode) plasmas. The pump-in is shown to be caused by a reduction of the turbulent transport and to be correlated with a change in the sign of the induced neoclassical transport. This novel reversal of the RMP induced transport has the potential to significantly improve reactor relevant, three-dimensional magnetic confinement scenarios.

The tokamak, benefiting from good confinement due to toroidal symmetry [1], is the leading candidate device for the magnetic confinement of burning plasma for energy production. Tokamaks can never be fully axisymmetric, however, and always have some level of 3D fields whether due to intrinsic asymmetries in the device construction or purposefully applied. Small, core resonant error fields (EFs, $\delta B/B_0 \approx 10^{-4}$) can destroy confinement by locking magnetic islands and must be corrected with applied resonant magnetic perturbations (RMPs) [2–9]. In so doing, overall asymmetry is often amplified through the non-resonant spectrum of fields (the spectrum not inducing core islands). Reactor relevant high confinement (H-mode, [10]) plasmas are subject to magnetohydrodynamic (MHD) instabilities called Edge Localized Modes (ELMs) that are also mitigated or suppressed by purposefully breaking toroidal symmetry with RMPs [11– 18]. Until now, it has been widely accepted that this breaking of the toroidal symmetry reduces particle confinement (referred to as “pump-out”). A reduction of 15-50% in confinement is common with the application of RMPs [12, 13, 16], due in large part to the formation of islands at the foot of the H-mode edge transport barrier or “pedestal” [19, 20]. This level of density pump-out is not necessary for core or edge stability, and much effort is being spent on to minimize the degradation through complicated quasi-symmetry optimizations [21] and real-time control techniques [22] in order to maximize fusion efficiency in the presence of RMPs. In contrast to these previous challenges, however, this report shows that RMPs naturally and robustly increase the particle confinement in certain rotation regimes of reactor relevant H-mode scenarios.

This work is unique in that it reports an observation of particle confinement improvement with the application of magnetic perturbations in H-mode plasmas. This is accompanied by a correspondingly novel reduc-

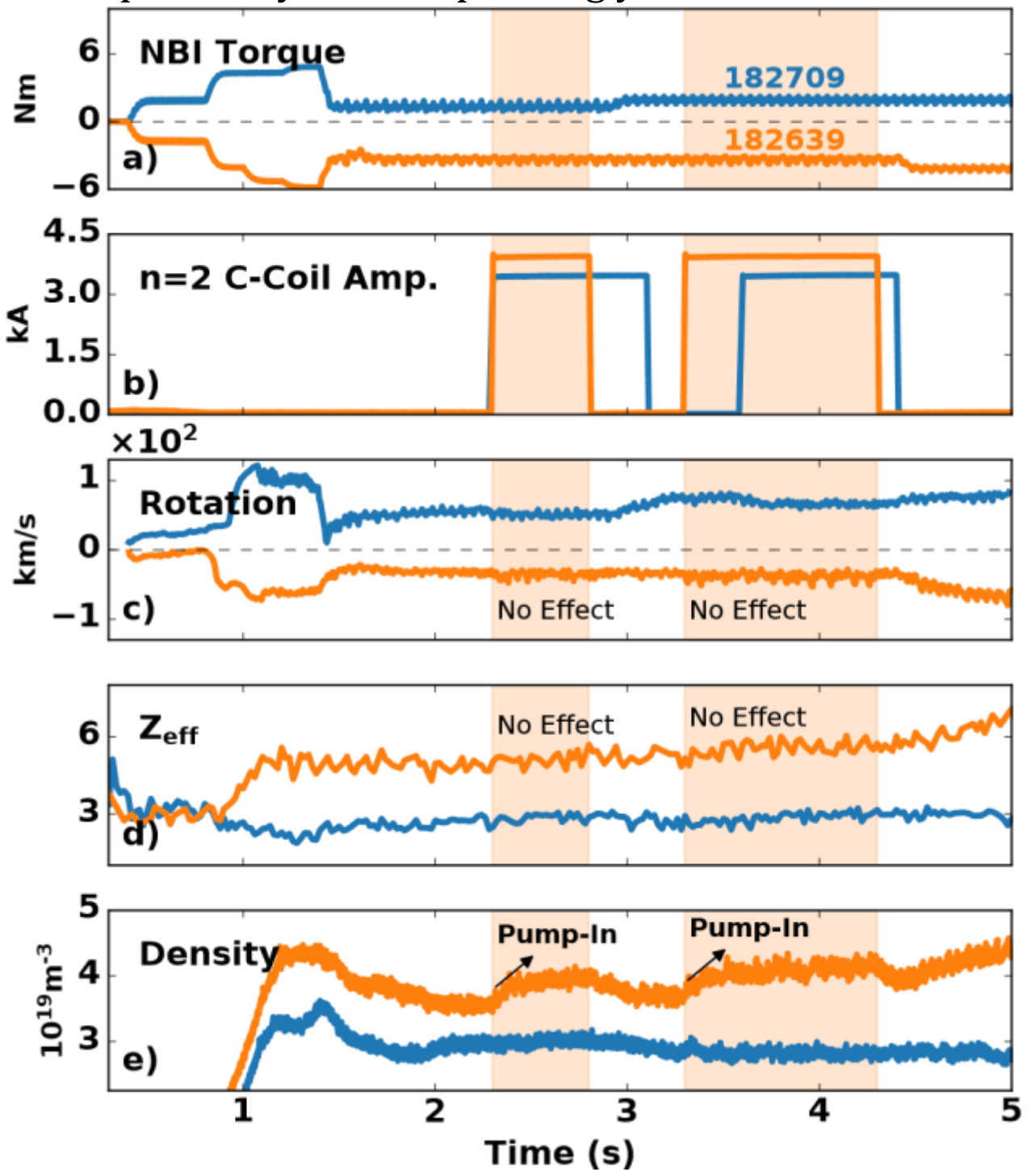

Figure 1. With constant NBI torque (a), the application of $n$ = 2 RMPs (b) in a discharge rotating (c) opposite Ip (orange) causes a sharp rise in electron density (d) without any correspondingly

sudden change in plasma composition (c). The effect is not present in co-Ip rotation (blue).

tion in edge turbulence with the application of RMPs. Past tokamak experiments have reported confinement improvements with applied non-axisymmetric fields following from changes to the plasma-wall interactions [23] and stability of large transients [24]. Confinement has also been shown to improve due to a sign change in trans-port across non-axisymmetric magnetic islands in certain rotation conditions [25]. The transport changes reported in this work, however, are independent of increased wall interactions that are not sustainable in a reactor and not reliant on large dynamic instabilities in the plasma. The reported confinement improvement with RMPs is thus the first applicable to tokamak reactor scenarios. It also represents a novel physics regime in which axial asymmetry actually improves confinement over the axisymmetric case, which has important implications for all magnetically confined plasmas (stellarators, in particular).

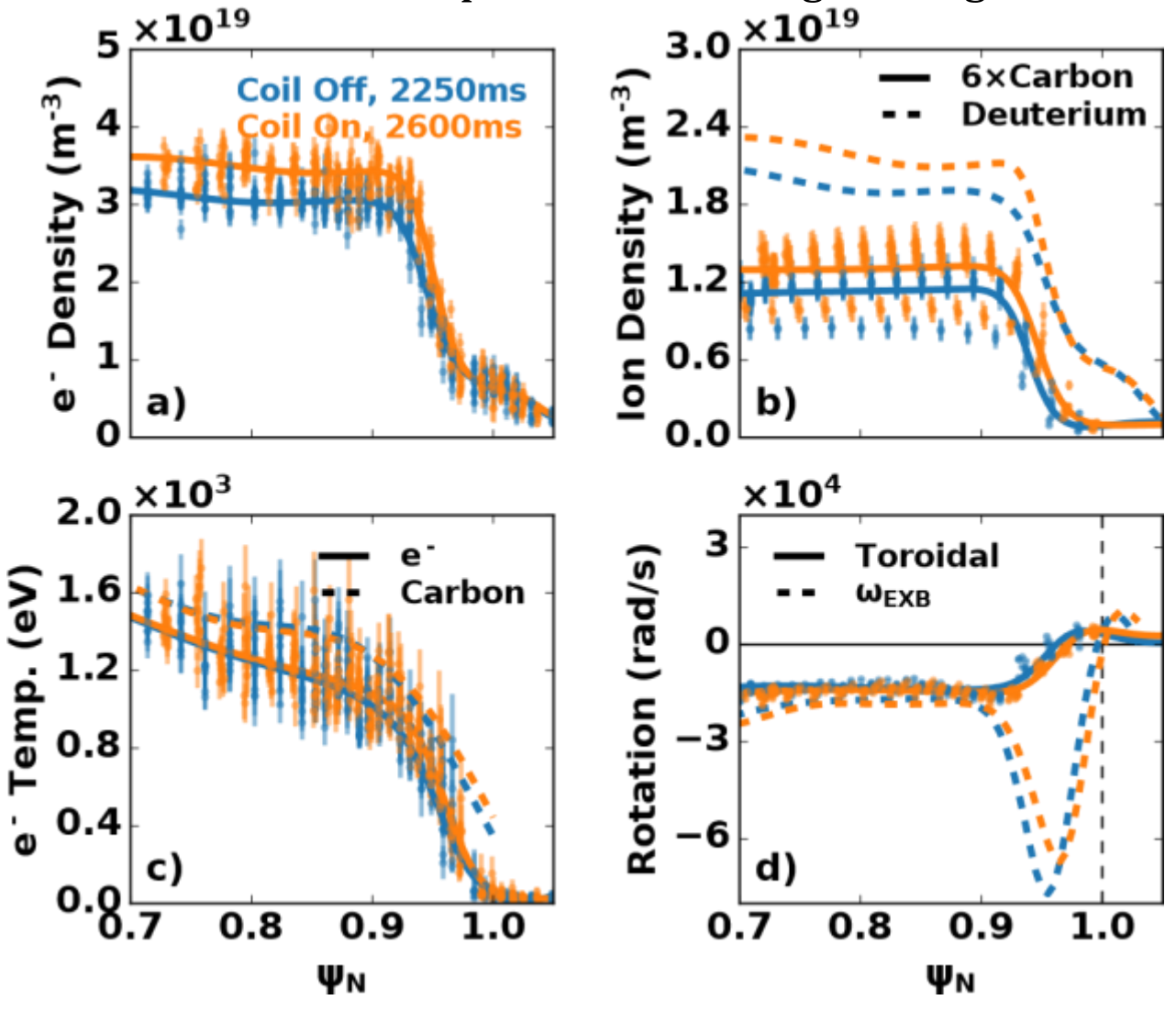

Figure 2. Measured electron density (a), ion density (b), temperature (c), and rotation frequency (d) pedestal profiles before (blue) and after (orange) the application of RMPs in counter-Ip rotating shot 182639. The densities rise while the temperatures and rotations are unaffected by the RMPs.

In recent DIII-D experiments, the application of RMPs consistently caused the density to rise in ELMing H-mode discharges across a range of moderate counter-Ip rotations. Ip is the toroidal plasma current such that counterIp is the ion diamagnetic drift direction. The ELMs in these plasmas are associated with proximity to the kinkpeeling stability boundary [26]. Experiments observing this phenomenon pulsed 3.4-5.7 kA, 310° $n = 2$ ($n$ is the toroidal harmonic) perturbations using DIII-D's midplane error field correction ("C-coil") array, which applies a mix of non-resonant and resonant magnetic perturbations [8, 27]. The amplitude of this RMP is below any (as yet undiscovered) ELM suppression threshold that may exist in these scenarios and above the 1 kA I-coil equivalent $n = 2$ intrinsic error field [28]. The confinement improvement is immediately observable in the line integrated density measured by an interferometer channel directed through the core of the plasma, as shown in figure 1. Detailed profile analysis in figure 2 shows the RMPs raise the pedestal density with little change to the temperature or rotation. The figure shows Thomson scattering (electron) and charge exchange recombination spectroscopy (carbon impurity assumed to be equilibrated with the main plasma deuterium ion temperature) measurement data as well as a lines indicating the Radial Basis Function fits from OMFIT [29, 30]. Quasineutrality has been assumed to calculate the main ion (deuterium) density and the profiles have been aligned to enforce the physical $\omega_{E\times B}(\psi_N = 1) = 0$ boundary constraint (the normalized poloidal flux $\psi_N$ is 1 at the plasma separatrix). These detailed profiles confirm the pump-in is in fact a confinement improvement impacting all species and not just an influx of impurities that increases the electron count.

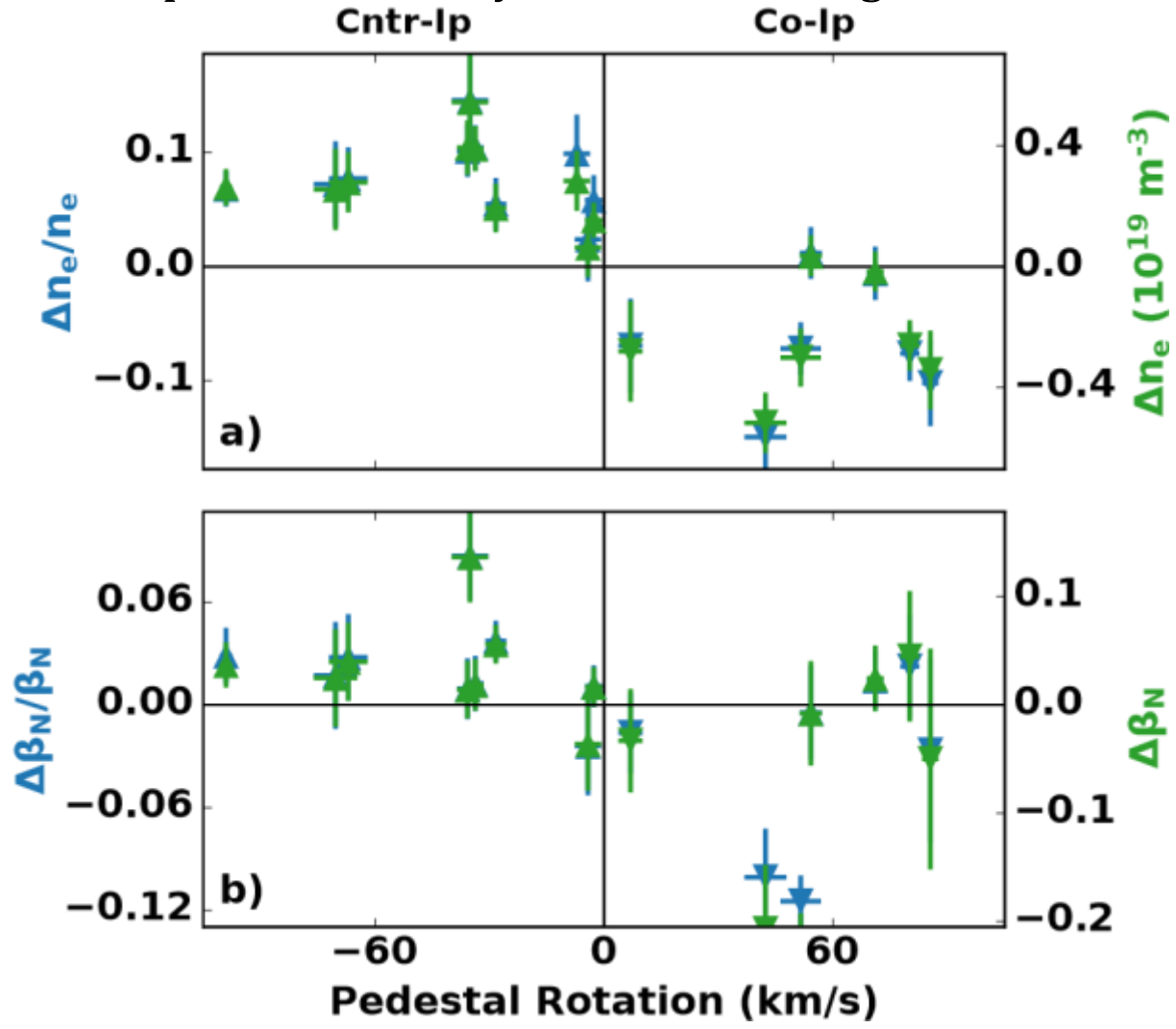

Figure 3. The change in line integrated density (a) and normalized pressure (b). The observed improvements (blue) and fractional changes (green) behave similarly, peaking at moderate negative rotation and reversing when the rotation becomes positive with respect to the plasma current. Upper single null (N) and lower single null (H) data is shown.

The neutral beam injected torque was scanned between discharges, and figure 3a shows the increase in density when applying RMPs peaks at almost 6 × $10^{18}$ $m^{-3}$, amounting to a 15% increase in the density. The improvement is reduced at the most counter-Ip rotations obtained in this experiment, the furthest of which is complicated by being a Quiescent H-mode (QH-mode, [31, 32]) with coherent edge MHD modes at countercurrent pedestal rotations below -80 km/s. Here, the pedestal rotation is taken from charge exchange measurements at $\psi_N \approx 0.89$. No gas fueling feedback was used during these shots and the line density (pedestal collisionality, $\nu_e^*$) naturally varies from 2.6 − 4.3 × $10^{19}$ $m^{-3}$(0.17 − 0.37) prior to RMP applications. These parameters do not separate the pump-in and pump-out and relative change overlays in figure 3 show these variations do not impact the observed rotation dependence. Figure 3b shows the density pump-in corresponds to a rise in normalized pressure that peaks with an observation of 13% improvement, consistent with the rise in density at roughly constant temperature. Figure 3 also includes data from typical lower single null ELM control target plasmas reported in Ref. [28] (distinguished by H markers) during the application of 4-4.6 kA $n = 2$ C-coil currents or 1.5 kA $n = 2$ currents in the internal I-coils (shown in [28] to have comparable resonant coupling to 4.6 kA of C-coil current). This data shows the usual density pump-out in positive rotations as well as hitherto unnoticed indications of pump-in at the slightly negative rotations obtained.

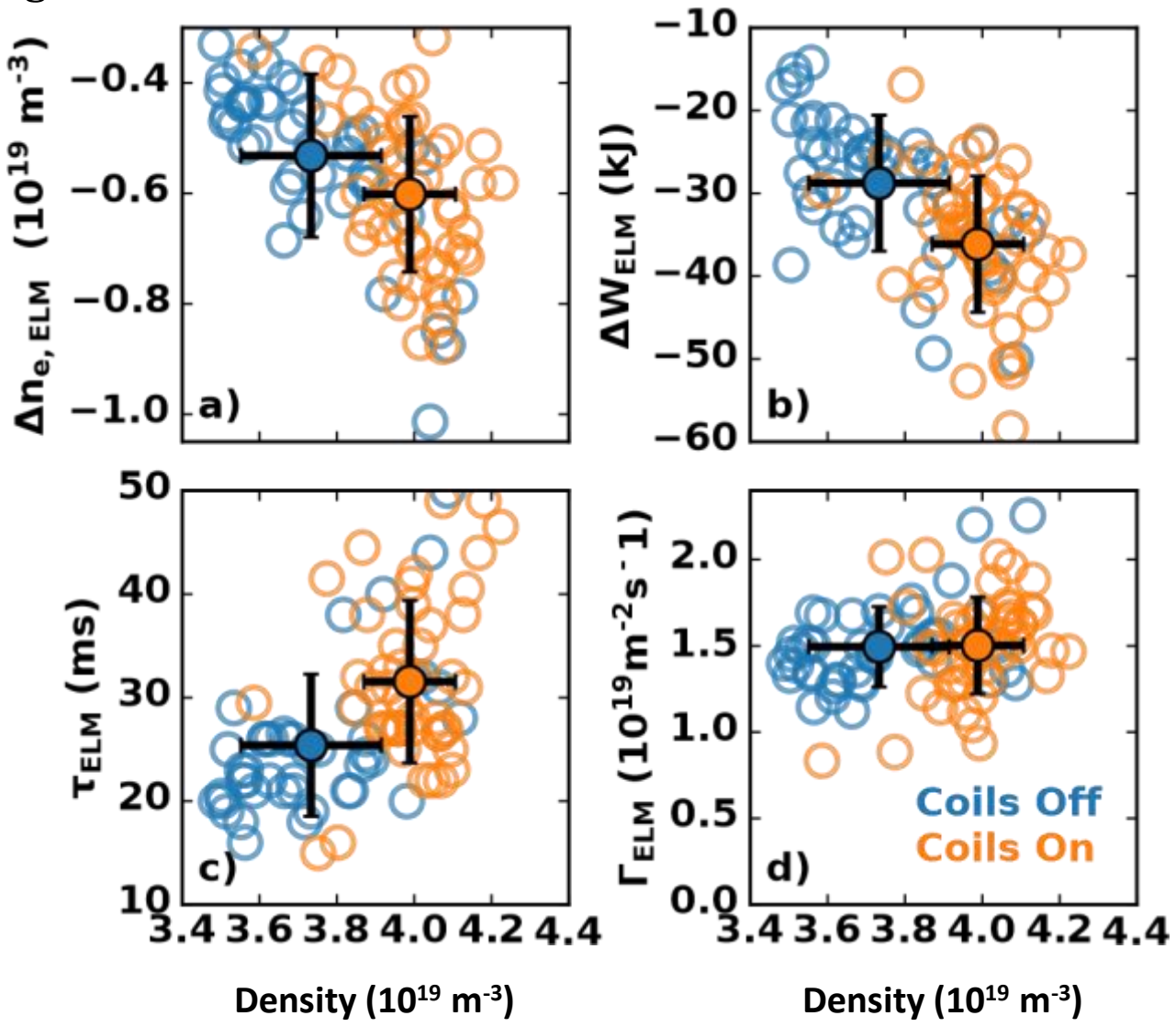

Figure 4. ELM characteristics for counter-Ip rotating shot 182639 without (blue) and with (orange) RMPs. Black error bars show the mean and standard deviation of each dataset. The RMPs increases the size of the ELM density (a) and energy (b) crashes but increase the ELM period (c), resulting in no change to the average ELM particle flux (d). The observed pump-in (separation of on/off points in the horizontal axis), thus does not come from a change in ELM particle flux.

The extensive suite of edge and plasma boundary diagnostics of DIII-D do not detect any changes in the wall particle source associated with the RMPs responsible for the observed rise in density. The exhaust rate, proportional to neutral pressure in the divertor region, is not impacted when the $n = 2$ RMP is applied. The Deuterium Balmer $\alpha$ ($D_\alpha$) emissivity from tangential camera views [33] (proportional to the deuterium density) show no qualitative changes with application of RMPs, indicating there is no change in particle sourcing. This is corroborated by the fact that $Z_{eff}$ (the effective charge state of the plasma) does not change at these times. Thus, we conclude the RMPs are modifying fundamental particle transport in the pedestal of these plasmas.

While no low-$n$ MHD exists to be impacted in the ELMing H-mode plasmas, figure 4 shows the ELMs become slightly larger with the application of the $n = 2$ RMP to these plasmas while their frequency decreases. The two effects effectively cancel when calculating the average ELM particle flux (the average rate at which ELM instabilities are expelling particles from the plasma). The difference in ELM fluxes with and without the RMP is 0.007 ± 0.365 × $10^{19}$ $m^{-2}s^{-1}$, with a value much smaller than the uncertainty. As the pump-in exceeds the statistical variance in the density during the coil-free phases, this rules out changes in the ELM behavior as the cause of the pump-in.

Note, the RMP pump-in causes ELM changes that are opposite of the widely observed "ELM mitigation" phenomenon, wherein RMPs applied below the threshold for ELM suppression result in smaller and more frequent ELMs [34–36]. It is, however, consistent with the known dependencies of ELM size and frequency with density (x-axis) [37]. The rise in density results in the change of ELM size and frequency, not the other way around.

The observed change in particle confinement is also distinct from the resonant island physics proposed for ELM suppression [15, 38] and particle pump-out [39, 40]. The pump-in plasmas have finite co-directional $E \times B$ and electron diamagnetic precession frequencies ($\omega_{E\times B}$ and $\omega_{*e}$ respectively) throughout the pedestal, shielding islands and providing no inward resonant transport across rational surfaces (which would require $\omega_{E\times B}/\omega_{*e} < -1$ in resistive MHD [25]). The absence of islands, determined by non-linear two-fluid MHD modeling using the TM1 code [39, 40], means the transport mechanism responsible for the pump-in does not need to overcome the common island induced pump-out.

Generalized Perturbed Equilibrium Code (GPEC, [41, 42]) calculations show the 3D-field-induced neoclassical non-ambipolar ion transport changes in the pedestal region are correlated with the observed changes in the particle confinement. Here, the pre-RMP profiles from figure 2 were used to form a kinetically constrained equilibrium from shot 182639 and the measured toroidal rotation was artificially scaled within the GPEC model to calculate it's impact on the neoclassical transport. The GPEC code is an equilibrium code and is not able to calculate the time dependent impact of this instantaneous flux. Simplified estimates of the modified density profiles using a constant effective diffusivity ($D_{eff}$) approximation, $n_{on} = -\int (\Gamma_{off} - \Gamma_{3D})/D_{eff} dr$, are shown in a black-red scale in figure 5b. The axisymmetric flux $\Gamma_{off}$ and effective diffusivity are calculated solving for power and particle balance using the ONETWO transport code [43] given the profiles from figure 2 prior to the RMP. The corresponding pedestal density change is shown in figure 5a and is small (0.2-1% change). It is always negative in this simple model because the edge $\omega_{E\times B}$, which is dominated by diamagnetic terms in the steep density gradient region, sets the sign of the midpedestal transport. Nonlinear studies have shown, however, that similar levels of the non-ambipolar neoclassical transport that changes sign at the top of the pedestal here can couple to the primary transport mechanisms existent in the symmetric state and result in the experimentally observed levels of density change [44–46]. While neoclassical model of torque reversals has been validated experimentally [47], this is the first observation of the particle transport reversing in a tokamak plasma and thus presents an opportunity for nonlinear modeling efforts to assess any possible contribution of this mechanism here.

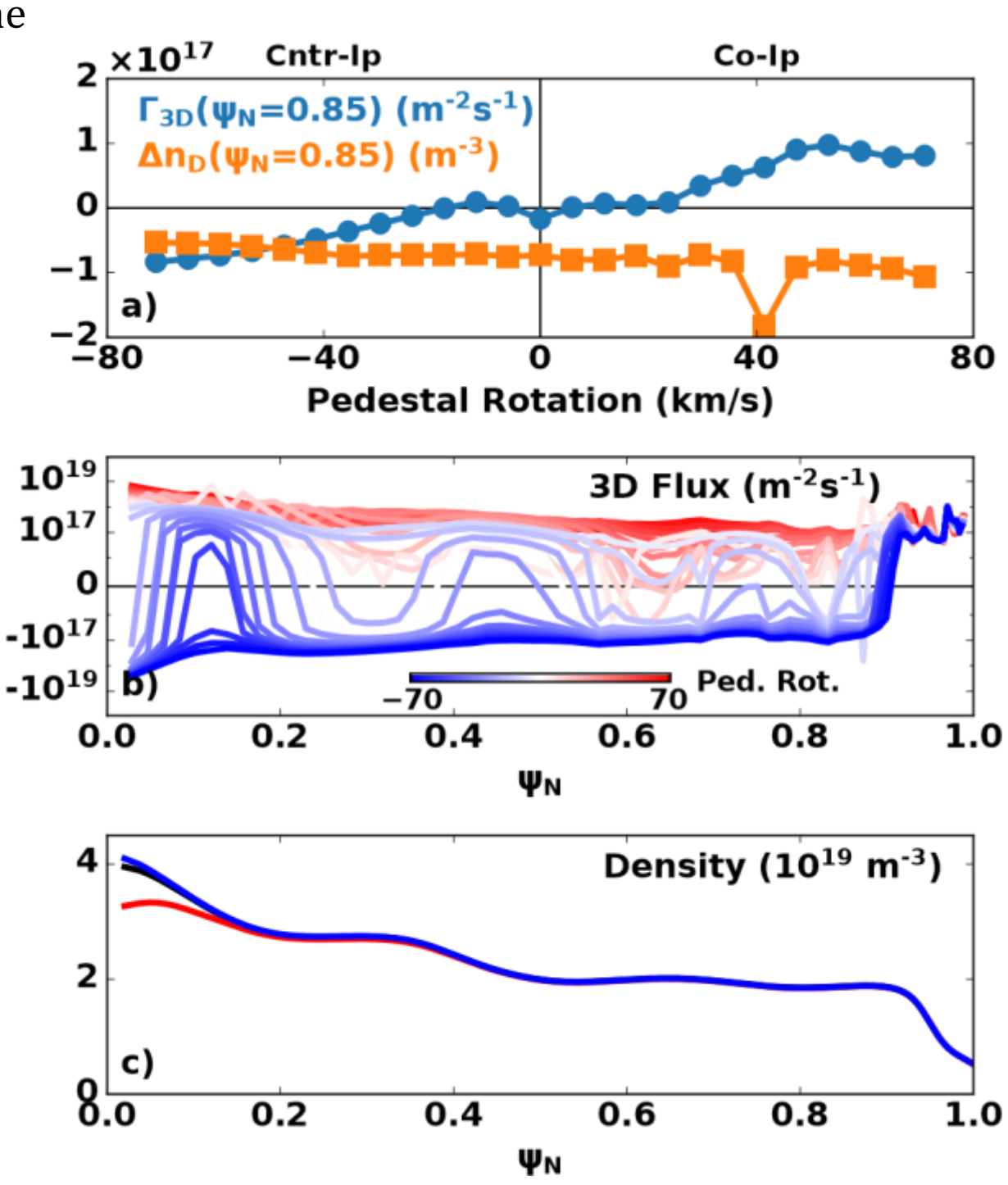

Figure 5. Magnetic perturbation induced, neoclassical ion transport across flux surfaces. Axes (a) shows the flux at the pedestal top ($\psi_N = 0.85$) changes for negative rotations but the linear estimate of pedestal pedestal density change does not. Axes (b) shows the neoclassical ion flux profiles and (c) shows corresponding linear estimates of the extreme case density profiles, which are relatively small deviations from the original (black) profile.

The nonlinear interplay between the 3D induced transport and existing transport mechanisms is observed experimentally in the mitigation of turbulent fluctuations coincident with the applied fields. Doppler back scattering (DBS) [48] analysis for a one of the peak pumpin pulses in figure 6b shows the density fluctuations with intermediate wavenumbers ($k_\perp = 4 - 6\mathrm{cm}^{-1}$ or $k_\perp \rho_s = 0.5 - 1.5$) change quickly with the application of the fields. The density then rises on a transport time scale within this suppressed turbulence state. The fast decrease in fluctuations, followed by the slower rise in density establishes a causal relationship between the RMP induced turbulence changes and the pump-in. A correspondingly sharp decrease in the measured phase velocity indicates that the decrease in the density fluctuation level is associated with a transition of the dominant inter-ELM turbulence from the ion-mode to the electronmode at the pedestal top, while the steep gradient region remains dominated by ion-mode turbulence. Note, this decrease in turbulence is opposite previous observations of turbulence enhancement with RMPs [36, 49, 50] and the underlying reason the turbulent transport is reduced is not yet known.

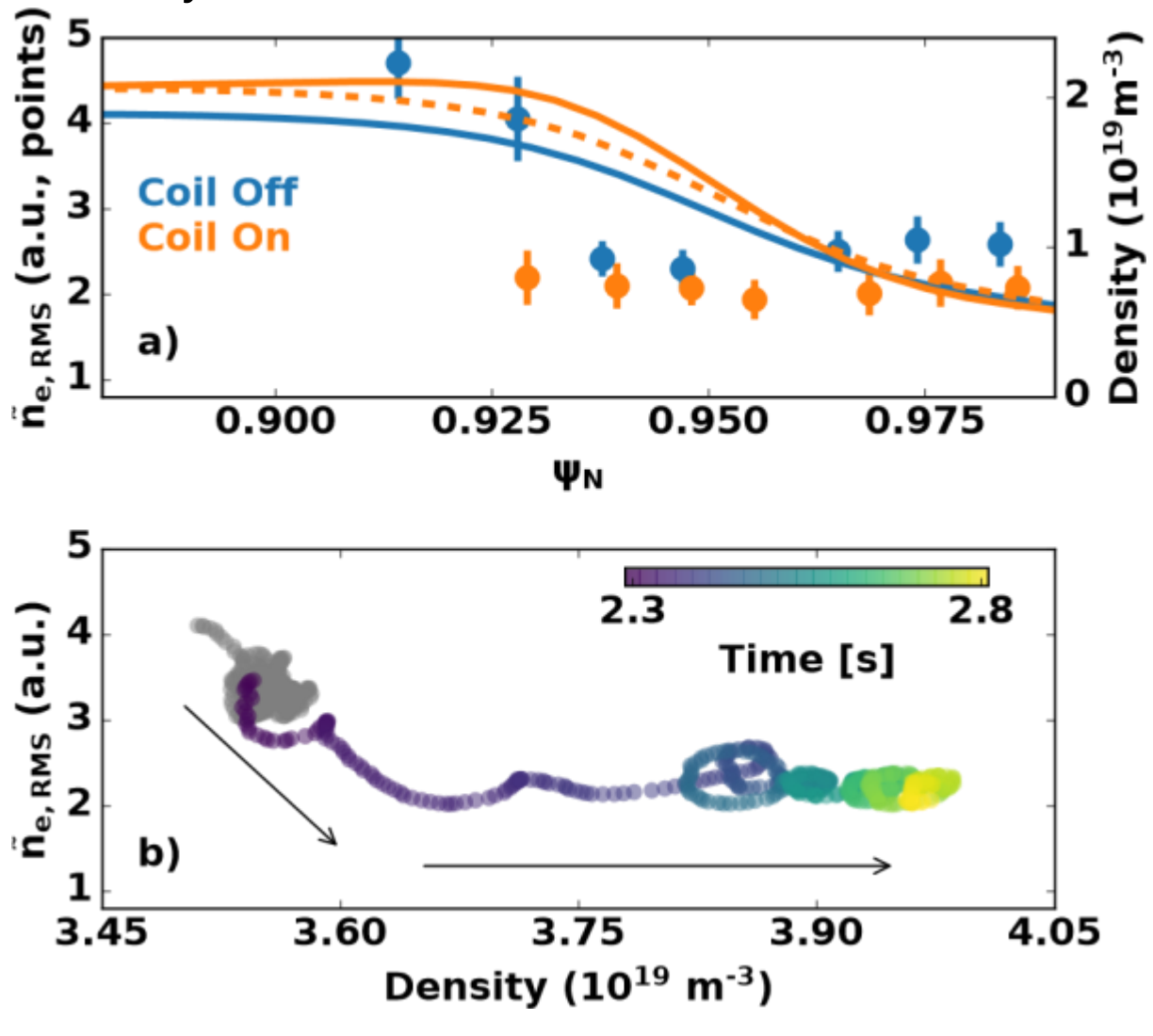

Figure 6. DBS density fluctuation measurements in discharge 182639. The fluctuations decrease across the entire pedestal when coils are applied (a), and has a causal influence on the observed density rise (b). The corresponding deuterium density model (a, dashed line) is comparable to the experimental rise in the profiles (a, solid lines).

The inter-ELM fluctuation profile measurements in figure 6a show this turbulence reduction is a robust feature across the entire pedestal in these scenarios. Previous DIII-D measurements show that this enhanced interELM edge turbulence measurement is highly correlated to increased particle transport [51]. A simple estimate of the ion density profile modification due to the change in turbulence is presented in figure 6a using the constant $D_{eff}$ approximation. Here, the 3D-induced flux profile is approximated assuming linear dependence on the turbulence amplitude $\Gamma_{3D} = \Gamma_{off}(A_{on} - A_{off})/A_{off} \approx 0.3\text{–}1\times10^{19}$ m$^{-2}$s$^{-1}$ in the outer radii where both amplitudes are available and assumed zero inside of this. Note, this is an order of magnitude larger than the modeled neoclassical particle flux and comparable to the $4 - 6 \times 10^{18}$ m$^{-2}$s$^{-1}$ values of edge flux computed from the experimental profiles using $dn/dt = \nabla \cdot \Gamma_{3D}$ during the pump-in. Accordingly, figure 6a shows the associated density change (calculated as in figure 5) corresponds to a rise in the deuterium pedestal density comparable to the one experimentally observed. A direct measurement of cross-field flux (currently not possible on DIII-D due to a lack of perturbed velocity measurements), is highly desirable for future qualitative studies of this phenomenon. It is clear, however, that this is an important mechanism for the observed pump-in.

In summary, DIII-D experiments have found a new regime in which RMPs like those planned for use in future H-mode reactors increase the particle confinement. A reduction in the turbulent particle transport with the application of RMPs is the dominant causal source of the pump-in. The neoclassical particle transport induced when breaking the toroidal symmetry also changes sign at the top of the pedestal, and should be modeled in more detail to determine the full extent of it's role (if any) in the observed pump-in. Both transport changes are distinct from the cross-island transport that causes pump-out in L and H-modes as well as any previously observed changed through sourcing or instabilities in Lmode plasmas. It is our hope that these new observations inspire 3D peeling mode model development to further understand the path to this stable rise in pedestal pressure. Future experimental work should test the relative diffusion and pinch transport terms in these regimes using gas puff modulation as well as test the compatibility between this confinement improvement and no ELM plasma regimes. While the pump-in regime dos not have the $\omega_{E\times B}$ zero crossing thought to be required for ELM suppression in DIII-D, the improvements are uniquely compatible with the use of RMPs in a reactor to correct error fields or support ELM free scenarios such as Quiescent H-modes.

The authors would like to thank Xiang Jian for his helpful discussions in the process of understanding the turbulent transport in these plasmas.

This work was supported by the U.S. Department of Energy, Office of Science, Office of Fusion Energy Sciences, using the DIIID National Fusion Facility, a DOE Office of Science user facility, under Awards DE-AC52-07NA27344, DE-AC02-09CH11466, DE-SC0019352, DE-SC0014264, DE-SC0022270, and DE-FC0204ER54698.

* logan35@llnl.gov

† Now at Columbia University

‡ Now at General Atomics

§ Now at Princeton Plasma Physics Laboratory